\documentclass[twocolumn,aps,pra]{revtex4}
\usepackage{epsfig}
\usepackage[english]{babel}
\usepackage{latexsym}
\usepackage{graphics}
\usepackage{subfigure}
\usepackage{epsfig}
\usepackage{graphicx}
\usepackage{dcolumn}
\usepackage{amsmath}
\usepackage{hyperref}
\usepackage{amssymb}

\begin{document}
\title{High ellipticity of harmonics from molecules in strong  laser fields of small ellipticity}

\author{F. J. Sun$^{1}$, C. Chen$^{1,2,*}$, W. Y. Li$^{3}$, X. Liu$^{4}$, W. Li$^{4}$, and  Y. J. Chen$^{1,\dag}$}
\date{\today}

\begin{abstract}

We study  high-order harmonic generation (HHG) from aligned molecules in strong elliptically polarized laser fields numerically and analytically.
Our simulations show that
the spectra and polarization of HHG depend strongly on the molecular
alignment and the laser ellipticity. In particular, for small laser ellipticity, large ellipticity of harmonics with high intensity
is observed for parallel alignment, with forming a striking ellipticity hump around the threshold.
We show that the interplay of the molecular structure and two-dimensional electron motion 
plays an important role here.
This phenomenon can be used to generate bright elliptically-polarized EUV pulses.

\end{abstract}
\affiliation{
1.College of Physics and Information Technology, Shaanxi Normal University,Xi'an710119, China\\
2.School of Physics, Hebei Normal University, Shijiazhuang 050024, China
\\3.School of Mathematics and Science, Hebei GEO University, Shijiazhuang 050031, China\\
4.Beijing Institute of Space Mechanics and Electricity, Chinese Academy of Space Technology, Beijing, 100094, China}

\maketitle

\section{Introduction}

In recent decades, high-order harmonic generation (HHG) from atoms \cite{Soc1,Huillier2,Tong3}, molecules \cite{Lein4,Itatani5,Corkum6} and solids \cite{Vampa7,Tao8,Dejean9}
has been a hot subject in experimental and theoretical studies of strong laser-matter interaction. The HHG has shown its promising
applications as a seed to generate attosecond  pulses. It can also be used as a unique tool to probe the ultrafast dynamics of electrons within atoms and molecules
with unprecedent attosecond resolution.

The HHG can be well understood with the classical \cite{Corkum10} and quantum \cite{Lewenstein11} three-step models.
These models describe the HHG as a three-step process:
 the valence electron of atoms or molecules is ionized by the laser field; the freed electron propagates in the external field;
the electron is driven by the laser field to return to and recombine with the parent ions with the emission of a high energy harmonic
along the direction parallel to the laser polarization (parallel harmonic).
The quantum model \cite{Lewenstein11}, frequently termed as strong-field approximation (SFA), also predicts the emission of harmonics
along the direction perpendicular to the laser polarization (perpendicular harmonic). For a linearly-polarized laser field and for linear symmetric molecules,
the perpendicular harmonics appear only for molecular targets aligned along a direction not parallel or perpendicular to the laser polarization \cite{Phan12}.
According to the SFA, parallel and perpendicular harmonics are emitted at the same instant and have no phase differences between them.
So the whole harmonics emitted are also linearly polarized.

Experimental studies indeed have observed the elliptical polarization of harmonics emitted by aligned molecules in a linearly polarized laser field \cite{Zhou13}.
This polarization effect of HHG has attracted great attentions in recent years,
as it implies that one can acquire elliptically-polarized XUV or EUV ultrashort pulses with HHG in a linearly-polarized laser field \cite{Levesque14}.
A great deal of experimental and theoretical efforts have been devoted to the complex origins of this HHG polarization \cite{Telnov15,Seideman16,Sherratt17,Strelkov18,Yu19,Dong20,li21}.
On the whole, this polarization is associated with the atomic or molecular properties in strong laser fields.
It occurs for the harmonics the parallel or perpendicular components of which are  subject to certain destructive interferences so that
different contributions are involved in the emission of these two components, resulting in a inherent phase difference between them.

When a two-dimensional (2D) laser field is used, such as a elliptically-polarized laser field \cite{Budil22,Dietrich23,Antoine24,Strelkov25,Strelkov26,Sarantseva27,Frolov28},
it is natural to think that the HHG from both atoms and molecules in such laser fields is also elliptically polarized.
However, in this case, how the molecular properties will affect this polarization is not very clear,
especially when the minor component of the 2D laser field is weak
so that it destroys the symmetry of the laser-driven system but does not have a remarkable influence on the electron dynamics.

In this paper, we focus on the polarization of HHG from aligned molecules \cite{Velotta29,Litvinyuk30} in strong elliptical laser fields with small ellipticity.
When the spectral properties of such HHG
have been studied widely \cite{Kanai31,Mairesse32,Odzak33,Yang34,Becker35},
the polarization properties of relevant HHG are less explored.
We pay attention to the issue if it is possible to obtain a bright and ultrashort pulse with large ellipticity using HHG polarization of molecules  in such laser fields.

Based on the numerical solution of the time-dependent Schr\"{o}dinger equation (TDSE) and SFA,
our simulations show that the polarization of HHG from  H$_2^+$  is strongly dependent on the molecular alignment and the laser ellipticity.
The HHG of the molecule shows the striking elliptical polarization located at some harmonic energy, and this location
shifts towards higher energy as increasing the alignment angle $\theta$
(the angle between the molecular axis and the main component of the elliptical laser field).
In particular, for parallel alignment of the molecule, for which the HHG elliptical polarization does not occur in a linearly-polarized laser field,
harmonics around the threshold show large ellipticity for a small laser ellipticity, with forming a striking ellipticity hump located at a wide energy region.
This phenomenon can be attributed to the interplay of the molecular structure and laser-induced 2D electron dynamics.
The small component of the elliptical laser field destroys the symmetry of the laser-driven system at $\theta=0^{\circ}$, allowing the emission of perpendicular harmonics at this angle.
At the same time, the parallel harmonics of $\theta=0^{\circ}$ around the threshold are subject to destructive intramolecular interference related
to the main component of the elliptical laser field, resulting in different phases of parallel and perpendicular harmonics.
This phenomenon  holds as we change the internuclear distance and the molecular species.
Because near-threshold harmonics including both parallel and perpendicular components generally have remarkably higher intensities
than those in the HHG plateau, the ellipticity hump for parallel alignment therefore can be used to obtain a strong elliptically-polarized EUV pulse.

\section{Theoretical methods}
\subsection{Numerical method}

We assume that the main component of the elliptical laser field is polarized along the direction parallel to the $x$ axis and the
minor one is along the $y$ axis. In addition, the molecular axis is located in the $xy$ plane.
Then the elliptical electric field can be written as  $\mathbf{E}(t)={\mathbf{e}}_{x}{E}_x(t)+{\mathbf{e}}_{y}{E}_y(t)$
with ${E}_x(t)=f(t)\mathcal{E}\sin{\omega_{0}}t$
and ${E}_y(t)=\varepsilon f(t)\mathcal{E}\sin({\omega_{0}}t+\phi)$ and $\phi=\pi/2$.
${\mathbf{e}}_{x}$  (${\mathbf{e}}_{y}$) is the unit vector along the $x$ ($y$) axis
(i.e., the major (minor) axis of the polarization ellipse).
 $\varepsilon$ is the laser ellipticity, $\omega_{0}$  is the laser frequency, and $f(t)$ is the envelope function.
$\mathcal{E}\equiv\mathcal{E}(\varepsilon)=\mathcal{E}_0/\sqrt{(1+\varepsilon^2)}$ and $\mathcal{E}_0$ is the laser
amplitude relating to the peak laser intensity $I$.

In the length gauge, the Hamiltonian of the molecule  interacting with the elliptical laser field can be written as
$H(t)={{{\textbf{p}}^{2}}}/{2}\;+\text{V}(\textbf{r})+\textbf{r}\cdot \textbf{E}(t)$
(in atomic units of $\hbar=e={{m}_{e}}=1$ ). Here, $\text{V}(\textbf{r})$ is the Coulomb potential of the molecule.
 For the H$^+_2$ system first explored in the paper, the  potential $\text{V}(\textbf{r})$ used has the form of
$V(\mathbf{r})=-Z/\sqrt{\zeta+\mathbf{r}_1^{2}}-Z/\sqrt{\zeta+\mathbf{r}_2^{2}}$
with $\mathbf{r}_{1(2)}^2=(x\pm \frac{R}{2}\cos\theta)^2+(y\pm \frac{R}{2}\sin\theta)^2$
in 2D cases. Here,
$\zeta=0.5$ is the smoothing parameter which is used to avoid the Coulomb singularity, and $\theta$ is the alignment angle.
$Z$ is the effective nuclear charge which is adjusted such that
the ionization energy of the model molecule at the internuclear distance $R$
is $I_p$=$1.1$ a.u.. Typically, for the equilibrium separation of $R=2$ a.u., we have $Z=1$.

Numerically, the TDSE of $i\dot{\Psi}(t)=H(t)\Psi(t)$ is solved with the spectral method \cite{Feit36}.
We use a grid size of $L_x \times L_y=204.8\times204.8$ a.u. with a grid spacing of $\Delta x=\Delta y=0.2$ a.u. for the $x$ and $y$ axis, respectively.
To eliminate the spurious reflections of the wave packet from the boundary, a mask function $F(r)$ is used in the
boundary to absorb the continuum wave packet. For the $x$ direction, we have used $F(x)={{\cos }^{{1}/{8}}}[\pi(|x|-x_{0})/(L_{
x}-2x_{0})]$ for $|x|\ge{{x}_{0}}$ and $F(x)=1$ for $|x|\textless{{x}_{0}}$. Here ${{x}_{0}}={{{L}_{x}}}/{8}\;$ is the boundary
 of the absorbing procedure. The situation is similar for the $y$ direction.
Alternatively, we can set the boundary of the absorbing procedure along the $x$ direction (i.e., the direction of the main polarization component of the elliptical laser field)
as $x_{0} = 0.9x_{m}$ with $y_{0} = L_y /8$ in the $y$ direction  unchanged  \cite{Yu19}.
Here, $x_m=\mathcal{E}/\omega_{0}^{2}$ is the maximal displacement of the electron as it travels in the laser field following the short trajectory.
In linearly-polarized cases, it has been shown that this treatment removes the contributions of the long trajectory and multiple returns to HHG,
but the short-trajectory contributions are not affected. For present elliptically-polarized cases with small ellipticity, we can also use the procedure to resolve short-trajectory contributions.
We will call this treatment short-trajectory simulations.
The interference between long and short trajectories has an important influence on harmonic polarization  \cite{Strelkov18}.
As discussed in  \cite{Yu19}, to identify the angle dependence of HHG polarization in a linearly-polarized laser field, the short-trajectory simulations are preferred.
This situation also holds in elliptical cases, as we will discuss below.

After the wave function $\Psi(t) $ is obtained,
the coherent parts of the parallel and perpendicular spectra can be evaluated with fourier transform of dipole acceleration:
\begin{equation}
\begin{aligned}
F_{\parallel(\perp)}(\omega) = \int\langle\Psi(t)|{\textbf{e}}_{x(y)}
\cdot\nabla V(\textbf{r})|\Psi(t)\rangle e^{i\omega t}dt.
\end{aligned}
\end{equation}
Here $\omega=n\omega_{0}$ is the emitted-photon frequency.
Then the intensities of the harmonic components can be written as $A_{\parallel(\perp)}(\omega)=\mid F_{\parallel(\perp)}(\omega)\mid^{2}$.

The ellipticity of HHG, which is related to the amplitude ratio and the phase difference of the parallel and perpendicular
harmonics, can be evaluated using

\begin{equation}
\begin{aligned}
\varepsilon_h=\sqrt{\frac{1+\eta-\sqrt{1+2\eta\cos2\delta+\eta^{2}}}{1+\eta+\sqrt{1+2\eta\cos2\delta+\eta^{2}}}},
\end{aligned}
\end{equation}
where $\delta=\phi_{\parallel}-\phi_{\perp}$ is the phase difference with $\phi_{\parallel(\perp)}(\omega)=arg[F_{\parallel(\perp)}(\omega)]$,
$\delta=k\pi+\delta_{1}$ with $k=0,1$ and $0\leq\delta_{1}\leq\pi$.
The term $\eta=A_{\perp}(\omega)/A_{\parallel}(\omega)$ denotes the amplitude ratio.
The subscript of the harmonic ellipticity $\varepsilon_h$ differentiates it from the laser ellipticity $\varepsilon$.
The range of harmonic ellipticity is $\varepsilon_h\in[0,1]$.
Equation (2) shows that high HHG ellipticity can be expected when the intensity of the perpendicular harmonic is
comparable to the parallel one and there is a phase difference of $\delta_{1}\sim\pi/2$ \cite{Dong20}.

In our simulations, we use a ten-cycle laser pulse which is linearly ramped up for two optical cycles and then kept at a constant intensity
for six additional cycles and finally linearly ramped down for two optical cycles.
Unless mentioned elsewhere, our discussions will be performed for the peak laser intensity of I$ = 5 \times 10^{14} \text{W/cm}^{2}$, the laser wavelength of $\lambda=800$ nm and $R=2$ a.u...

\subsection{Analytical description}
In analytical treatments, we assume that the molecular axis is along the $z$ axis, and the  laser field
$\textbf{E}(t)$ is located in the $xz$ plane with an angle $\theta$
between its major axis  and the molecular axis.
Then according to the SFA, the time-dependent dipole moment can be written as \cite{Lewenstein11}
\begin{equation}
\begin{aligned}
\textbf{x}(t)= &i\int_{0}^{\infty}d\tau \big[\xi(\tau)\textbf{d}_{r}^{*}(\textbf{p}_{st}+\textbf{A}(t))e^{-iS(\textbf{p}_{st},t,\tau)}\\ & \times\textbf{E}(t-\tau)\cdot
\textbf{d}_{i}(\textbf{p}_{st}+\textbf{A}(t-\tau))\big]+c.c. .\\
\end{aligned}
\end{equation}
Here, $\tau=t-t'$ is the excursion time of the rescattering electron in the driving laser field when it is ionized at the time $t'$, and
$\xi(\tau)=(\frac{\pi}{\epsilon'+i\frac{\tau}{2}})^{\frac{3}{2}}$ with infinitesimal $\epsilon'$.
The term $\textbf{p}_{st}\equiv \textbf{p}_{st}(t,t')=-\frac{1}{t-t'}\int^{t}_{t'}\textbf{A}(t'')dt''$ is the canonical
momentum, and $\textbf{A}(t)=-\int^{t}\textbf{E}(t')dt'$  is
the vector potential of the external field $\textbf{E}(t)$.
The term $S(\textbf{p}_{st},t,\tau)=\int^{t}_{t'}dt''{\{\frac{[\textbf{p}_{st}+\textbf{A}(t'')]^2}{2}+I_{p}\}}$ is the semiclassical action. 
The term $\textbf{d}_{i}(\textbf{p})$ is the bound-free dipole transition matrix element between the molecular ground state $|0\rangle$ and
the continuum $|\textbf{p}\rangle$ (which is approximated with the plane wave $|e^{i\textbf{p}\cdot\textbf{r}}\rangle$) in the ionization step
and can be written as
$\textbf{d}_{i}(\textbf{p})=\langle\textbf{p}|\textbf{r}|0\rangle=(2\pi)^{-3/2}\int d\textbf{r} e^{-i\textbf{p}\cdot\textbf{r}}\textbf{r}\langle\textbf{r}|0\rangle$.
Similarly, the term $\textbf{d}_{r}(\textbf{p})=\langle\textbf{p}_{k}|\textbf{r}|0\rangle=(2\pi)^{-3/2}\int d\textbf{r} e^{-i\textbf{p}_{k}\cdot\textbf{r}}\textbf{r}\langle\textbf{r}|0\rangle$
is that in the recombination step with the effective momentum $\textbf{p}_{k}$ which considers the Coulomb correction
on the momentum $\textbf{p}$ of the continuum $|\textbf{p}\rangle\propto|e^{i\textbf{p}\cdot\textbf{r}}\rangle$ in recombination.
Note, this correction changes only the momentum  $\textbf{p}$ of the state $|\textbf{p}\rangle$,
with assuming the energy $E_p=\textbf{p}^2/2$ of the state $|\textbf{p}\rangle$ unchanged. For linearly-polarized cases, one can use the expression of $\textbf{p}_{k}=\frac{\textbf{p}}{|\textbf{p}|}p_k$ with ${p}_{k}=\sqrt{2(E_p+I_p)}$  \cite{Chen37}.
We will discuss the form of the effective momentum $\textbf{p}_{k}$ for the present elliptically-polarized cases with small laser ellipticity later.

Through fourier transform of $\textbf{x}(t)$, the coherent part of the spectrum along the major axis $\textbf{e}_{\parallel}$ of the elliptical laser field  can be written as
\begin{equation}
\begin{aligned}
{F}_{l}(\omega)=& i\int dt \int_{0}^{\infty}d\tau\big[\xi(\tau)\textbf{e}_{\parallel}\cdot \textbf{d}_{r}^{*}(\textbf{p}_{st}+\textbf{A}(t))\\ &\times \mathbf{E}(t-\tau)\cdot
\textbf{d}_{i}(\textbf{p}_{st}+\textbf{A}(t-\tau))e^{-iS(p_{st},t,\tau)} e^{i\omega t}\big].
\end{aligned}
\end{equation}
The integration in the above expression can be treated with solving the saddle-point equation \cite{Lewenstein11,Salieres38}
\begin{equation}
\begin{aligned}
&[\textbf{p}_{st}+\textbf{A}(t'_{s})]^{2}/2+I_{p}=0;\\
&[\textbf{p}_{st}+\textbf{A}(t_{s})]^{2}/2+I_{p}=\omega.
\end{aligned}
\end{equation}
The first equation describes the tunneling process with the ionization momentum $\textbf{p}_{si}=\textbf{p}_{st}+\textbf{A}(t'_{s})$, 
and the second equation describes the recombination process with the recollision momentum $\textbf{p}_{sr}=\textbf{p}_{st}+\textbf{A}(t_{s})$. 
These momenta generally have complex forms in 2D laser fields.
Solving the saddle-point equations, one can get the saddle-point ionization time $t'_{s}$ and saddle-point return time $t_{s}$ of the
rescattering electron, as well as the saddle-point momentum $\textbf{p}_{st}(t_{s},t'_{s})$.
So Eq. (4) can be simplified as
\begin{equation}
\begin{aligned}
{F}_{l}(\omega)\propto&
\textbf{}\sum_s[G(t_{s},\tau_{s})\textbf{e}_{\parallel}\cdot \textbf{d}_{r}^{*}(\textbf{p}_{sr}) 
\textbf{E}(t'_{s})\cdot \textbf{d}_{i}(\textbf{p}_{si})S_{p}(\omega)].
\end{aligned}
\end{equation}
Here, $G(t_{s},\tau_{s})=\xi(\tau_{s})[1/det(t_{s},\tau_{s})]^{1/2}$, and
$det(t_{s},\tau_{s})$ denotes the determinant of the $2 \times 2$ matrix formed by the second derivatives of the action with respect to
$t$ and $\tau$ \cite{Lewenstein39}.
The term  $\tau_{s}=t_{s}-t'_{s}$ is the saddle-point travel time.
The sum in Eq. (6) extends over all possible saddle points $(t'_{s},t_{s})$ for the emission of a harmonic
 $\omega$. The  term $S_{p}(\omega)$ has the form of
$S_{p}(\omega)=e^{-i(S_{s}-\omega t_{s})}$, with $S_{s}\equiv S(\textbf{p}_{st},t_{s},\tau_{s})$.
The real parts of the saddle points $(t'_{s},t_{s})$ have been considered as the ionization and return times of the electron, respectively.
The saddle points have also been termed as electron trajectories including long trajectory, short trajectory and multiple returns.
The trajectories are well resolved in the temporal region and have different ionization and return times.

In Eq. (6), these two dipoles $\textbf{e}_{s}\cdot \textbf{d}_{i}(\textbf{p}_{si})$ and $\textbf{e}_{\parallel}\cdot \textbf{d}_{r}^{*}(\textbf{p}_{sr})$
are mainly responsible for the angle dependence of HHG, as discussed in \cite{Chen40}.
The symbol $\textbf{e}_{s}$ denotes the unit vector along the laser polarization of the electric field $\textbf{E}(t)$ at $t=t'_{s}$.
We write the product of these two dipoles at the saddle point $(t_{s},t'_{s})$ as
\begin{equation}
\begin{aligned}
M_{s}(\omega,\theta)=&|\textbf{e}_{s}\cdot \textbf{d}_{i}(\textbf{p}_{si})|^{2}
\cdot|\textbf{e}_{\parallel}\cdot \textbf{d}_{r}^{*}(\textbf{p}_{sr})|^{2}.
\end{aligned}
\end{equation}
With describing the ground-state wavefunction using linear combination of atomic orbitals-molecular orbitals (LCAO-MO) approximation,
the dipole moment for  H$_{2}^{+}$ with $1\sigma_{g}$ valence orbital  can be written as \cite{Chen37}
\begin{equation}
\begin{aligned}
\textbf{d}^{1\sigma_{g}}(\textbf{p})= N^{1\sigma_{g}}[-2i\cos(\frac{\textbf{p}\cdot \textbf{R}}{2})\cdot \textbf{d}^{1s}(\textbf{p})].
\end{aligned}
\end{equation}
Here,  $N^{1\sigma_{g}}$ is the normalization factor, and $\textbf{R}$ is the vector between the two atomic cores of the molecule.
The term $\cos(\textbf{p}\cdot\textbf{R}/2)$  denotes interference between these two cores and $\textbf{d}^{1s}(\textbf{p})$ denotes the atomic dipole
moment of $1s$ orbital. When the laser ellipticity $\varepsilon$ is small so that the main component of the elliptical laser field dominates ionization,
the ionization momentum ${p}_{si}$ can be approximately expressed with ${p}_{si}\approx\pm i\kappa=\pm i\sqrt{2I_p}$ \cite{Xie41},
then Eq. (7) can be rewritten as \cite{Chen40,Chen37,Chen42,Li43}
\begin{equation}
\begin{aligned}
\noindent
M_{s}(\omega,\theta)\propto \mid \cos(i\kappa \frac{R}{2}\cos{\theta_I})\mid^{2}M^r_{s}(\omega,\theta)M_s^a,
\end{aligned}
\end{equation}
with
\begin{equation}
\begin{aligned}
\noindent
M^r_{s}(\omega,\theta)=\mid\cos({p}'_{sr}\frac{{R}}{2}\cos\theta_R)\mid^{2}.
\end{aligned}
\end{equation}
Here, $M_s^a\equiv\mid(\textbf{e}_{s}\cdot \textbf{d}_i^{1s}(\textbf{p}_{si}))(\textbf{e}_{\parallel}\cdot \textbf{d}_r^{1s*}(\textbf{p}'_{sr}))\mid^{2}$,
and $\theta_I$ ($\theta_R$) is the exit (recollision) angle  between the vectors  $\textbf{p}_{si}$ ($\textbf{p}_{sr}$) and  $\textbf{R}$ \cite{Chen44}. In Eq. (10), we have used the effective momentum $\textbf{p}'_{sr}$ instead of $\textbf{p}_{sr}$ in the recombination  dipole to consider the Coulomb correction, as discussed in Eq. (3).
For the elliptical laser field with a small ellipticity $\varepsilon$ explored here,
the effective momentum $\textbf{p}'_{sr}$ used has the form of
$\textbf{p}'_{sr}=\textbf{e}_{\parallel}p'_{\parallel}+\textbf{e}_{\perp}p_{\perp}$ with ${p}'_{sr}=\sqrt{p'^2_{\parallel}+p^2_{\perp}}$.
Here, the symbol $\textbf{e}_{\perp}$ denotes the unit vector along
the minor axis of the elliptical laser field, $p'_{\parallel}=\sqrt{2[\omega-\varepsilon I_{p}]}$ and
$p_{\perp}$ being the real part of the $\textbf{e}_{\perp}$ component of $\textbf{p}_{sr}=\textbf{p}_{st}+\textbf{A}(t_{s})$
at the saddle point $(t'_{s},t_{s})$. Accordingly, the recollision angle $\theta_R$ has the expression of
$\theta_R=\arctan(p_{\perp}/p'_{\parallel})-\theta$.

As discussed in  \cite{Chen37}, for H$_{2}^{+}$ in linearly polarized laser fields,
the term $M^r_{s}(\omega,\theta)$ in Eq. (9) is the most  sensitive to the molecular alignment.
Our simulations show that the situation also holds for the cases of small laser ellipticity in the paper.
In the following, we will compare the predictions of $M^r_{s}(\omega,\theta)$ of Eq. (10) with the TDSE results to understand the angle-dependent HHG ellipticity.

\begin{figure}[t]
\begin{center}
\rotatebox{0}{\resizebox *{8.5cm}{6.5cm} {\includegraphics {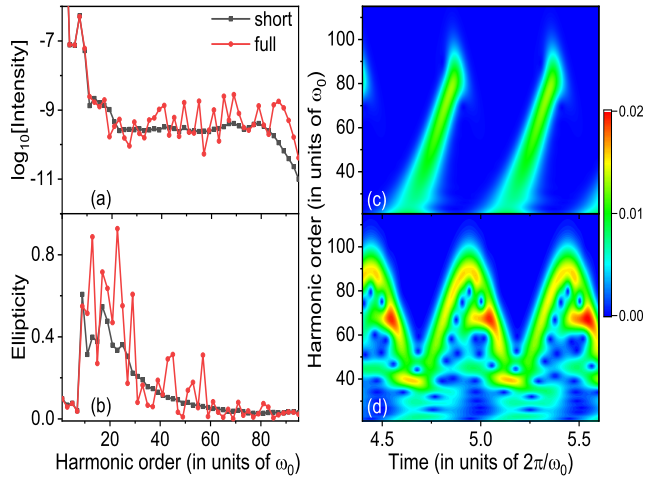}}}
\end{center}
\caption{Spectra (a) and  ellipticity (b) of harmonics  for aligned molecules H$_{2}^{+}$ at $\theta=0^{\circ}$ and $\varepsilon=0.1$,
obtained with short-trajectory (black square) and full (red circle) TDSE simulations. In (c) and (d), we show relevant
time-frequency distributions for short-trajectory (c) and full (d) simulations.
} \label{Fig. 1}
\end{figure}

\section{Results and Discussions}

In Fig. 1, we show the  HHG results of full TDSE simulations and short-trajectory simulations at $\varepsilon=0.1$ and $\theta=0^{\circ}$.
As the spectra of full simulations show the complex interference structure, the short-trajectory one is more smoothing, as seen in Fig. 1(a).
Accordingly, the ellipticity of short-trajectory harmonics is also more regular than the full-simulation one, with showing a remarkable ellipticity hump around the threshold, as seen in Fig. 1(b).
This ellipticity hump at $\theta=0^{\circ}$ which disappears in a linearly-polarized laser field is the main phenomenon we will discuss in the paper.
In Fig. 1(d). the  time-frequency analysis of TDSE dipole acceleration \cite{Tong45} also clearly shows the complex interference between different electron trajectories in full TDSE simulations,
as the interference structure is basically absent in  short-trajectory results in Fig. 1(c).
In the following, we focus on the short-trajectory results which allow a clear identification of the angle-dependence polarization of HHG from molecules in the elliptical laser field.

\begin{figure}[t]
\begin{center}
\rotatebox{0}{\resizebox *{8.5cm}{8.5cm} {\includegraphics {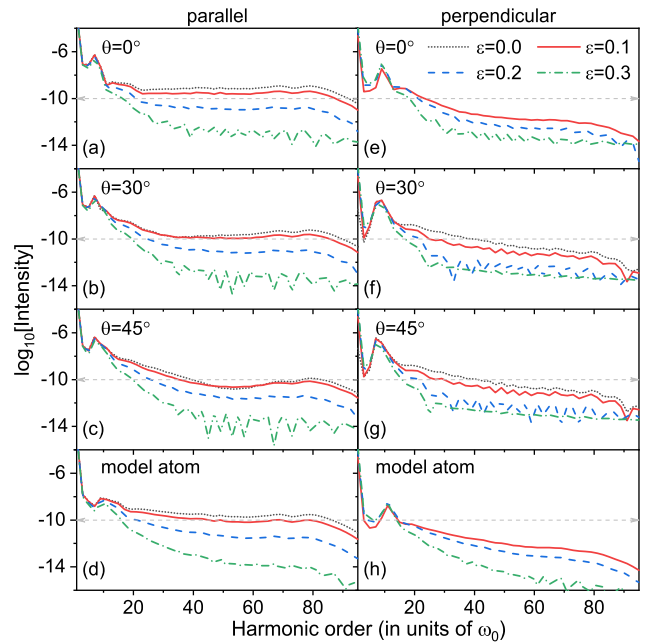}}}
\end{center}
\caption{Spectra of parallel (a-d) and perpendicular (e-h) harmonics for aligned molecules H$_{2}^{+}$ at  $\theta=0^{\circ}$ (a,e),
$30^{\circ}$ (b,f), $45^{\circ}$ (c,g) and for model atom with similar $I_p$ to H$_{2}^{+}$ (d,h), with the laser ellipticity of  $\varepsilon=0$ (black dotted),
$\varepsilon=0.1$ (red solid), $\varepsilon=0.2$ (blue dashed) and $\varepsilon=0.3$ (green dashed-dotted).
The horizontal arrows in each row are plotted to facilitate the comparison.
} \label{Fig. 2}
\end{figure}

To understand the polarization of HHG from  aligned molecules in elliptical laser fields, in Fig. 2, we plot the TDSE short-trajectory spectra of
parallel versus perpendicular harmonics of H$_2^+$ at different angles $\theta$ and laser ellipticity $\varepsilon$. For comparison, the results of a model atom with $I_p=1.1$ a.u. are also shown.

For cases of parallel harmonics in the left column of Fig. 2, the results in each panel show that the intensities of spectra decrease fast when the laser ellipticity is increased,
reflecting the suppression of the recombination due to the lateral motion of the electron induced by the minor component of the elliptical laser field.
For cases of perpendicular harmonics, the intensities of spectra also decrease with the increase of laser ellipticity on the whole, as seen in each panel in the right column of Fig. 2.
It is worth noting that for $\theta=0^{\circ}$ in Fig. 2(e) and for the atom case in Fig. 2(h), the perpendicular harmonics disappear for $\varepsilon=0$ corresponding to a linearly-polarized laser field,
due to the symmetry of the laser-driven system in the cases \cite{Phan12}.
In addition, results in Fig. 2 also show that for lower harmonic orders below the threshold (about H19), the yields of parallel and perpendicular harmonics
are remarkably higher than those in the plateau regions of the spectra and are not sensitive when changing the laser ellipticity.
In particular, they are comparable  at small laser ellipticity, suggesting the possibility of a high harmonic ellipticity according to Eq. (2).
With the help of the horizontal arrows in each panel, a careful comparison for the molecular cases also shows that both of the parallel and perpendicular spectra differ for the molecular alignment and
the parallel spectra in the plateau region  are stronger  at $\theta=0^{\circ}$ than other cases. We will return to this point later.

\begin{figure}[t]
\begin{center}
\rotatebox{0}{\resizebox *{8.5cm}{8.5cm} {\includegraphics {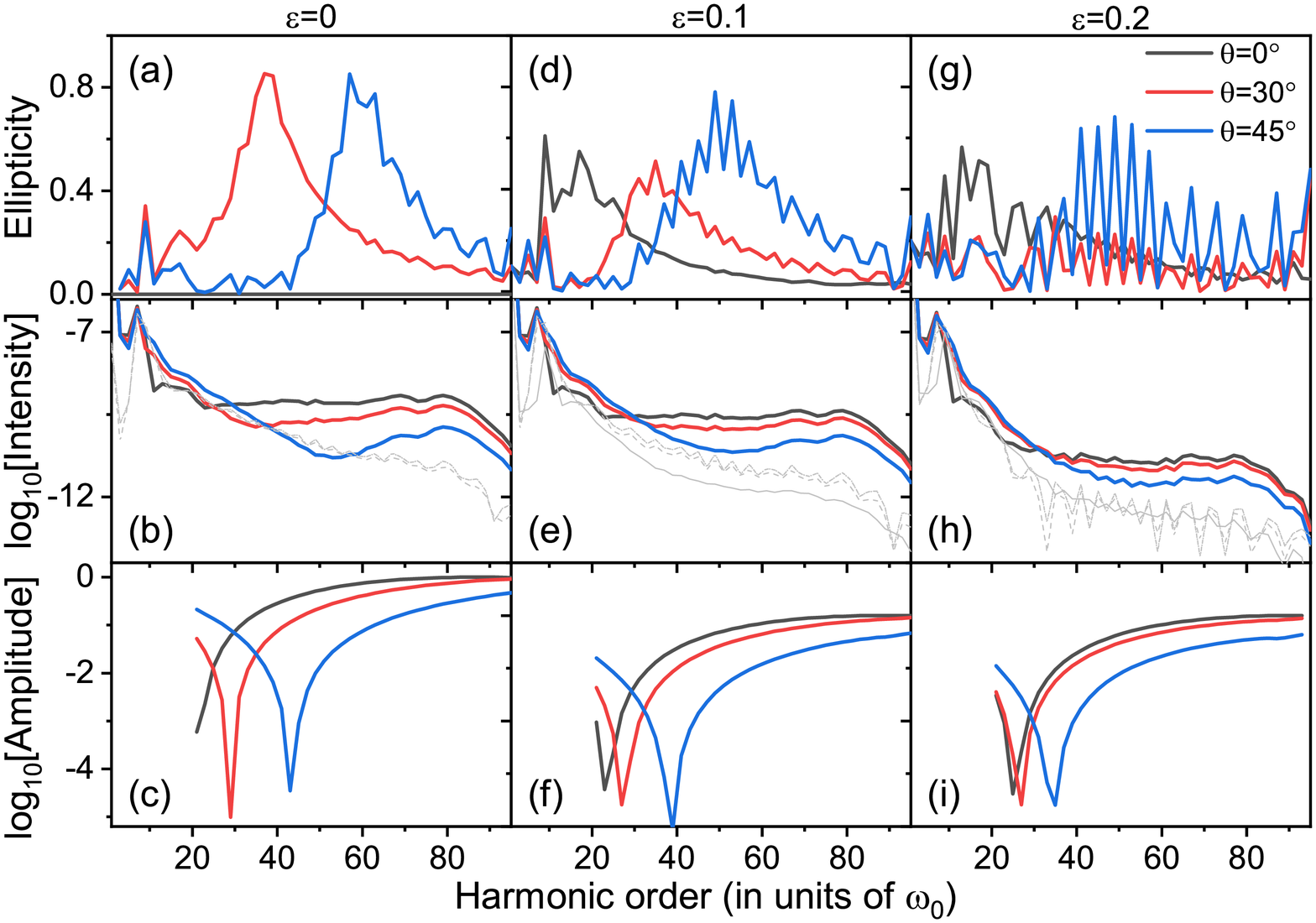}}}
\end{center}
\caption{Ellipticity (the first row), spectra of parallel harmonics (second) and the corresponding function curves $M^r_{s}(\omega,\theta)$ of Eq. (10) (third)
for aligned molecules H$_{2}^{+}$ at different  angles of
$\theta=0^{\circ}$ (black curves), $\theta=30^{\circ}$ (red), and $\theta=45^{\circ}$ (blue) and different laser ellipticity  of $\varepsilon=0$ (the first column),
$\varepsilon=0.1$ (second), and $\varepsilon=0.2$ (third). In the second row, spectra of perpendicular harmonics at $\theta=0^{\circ}$ (solid),
$\theta=30^{\circ}$ (dashed) and $\theta=45^{\circ}$ (dotted) are also plotted with gray curves for comparison.
} \label{Fig. 3}
\end{figure}

In Fig. 2, our discussions mainly concentrate on the dependence of HHG yields of parallel versus perpendicular harmonics
on the laser ellipticity for a certain alignment angle.
In Fig. 3, we present the comparison between HHG ellipticity, the corresponding spectra and dipoles of Eq. (10)
at different angles for a certain laser ellipticity. Firstly, for  $\varepsilon=0$ of the linearly-polarized case in Fig. 3(a),
harmonics at $\theta=30^{\circ}$ and $\theta=45^{\circ}$ show large ellipticity, with forming an ellipticity peak around some harmonic energy.
The position of the ellipticity peak shifts with the increase of the angle and corresponds to the minimum in the relevant parallel spectrum in Fig. 3(b).
These spectral minima are associated with two-center interference and are well described by the corresponding minima in the function curves of Eq. (10)
 in Fig. 3(c).  For simplicity, we will call the function curve  of Eq. (10) the dipole below.
A careful comparison tells that the spectrum at $\theta=0^{\circ}$ in Fig. 3(b) also shows a minimum at about H21 which is near the threshold,
in agreement with the prediction of $M^r_{s}(\omega,\theta)$ in Fig. 3(c).
However, the harmonics at this angle do not show the ellipticity due to the absence of perpendicular harmonics.
For the case of a small laser ellipticity of  $\varepsilon=0.1$ in Fig. 3(d),
a striking ellipticity plateau of harmonics located around the threshold
is observed for $\theta=0^{\circ}$, which has been indicated in Fig. 1.
For this laser ellipticity in Fig. 3(d), harmonics at $\theta=30^{\circ}$ and $\theta=45^{\circ}$  also show the remarkable ellipticity peak,
the magnitudes of which are somewhat smaller than the corresponding ones in Fig. 3(a). When comparing with the spectra in Fig. 3(e),
one can also observe that large ellipticity of harmonics generally appears at the harmonic orders at which
the harmonic spectrum shows a striking minimum, as discussed in the first column of Fig. 3. The positions of the spectral minima in Fig. 3(e)
also basically agree with the positions of the dipole minima, as shown in Fig. 3(f). In comparison with the results in Fig. 3(c), one can also observe from Fig. 3(f)
that the minima in dipoles for  intermediate angles of $\theta=30^{\circ}$ and $\theta=45^{\circ}$ shift somewhat towards lower energy, while
the minimum in the dipole of $\theta=0^{\circ}$ shifts somewhat towards higher energy. This is due to the influence of the minor component of the elliptical laser field,
which changes the recollision  angle of HHG. We will address the question in detail in Fig. 4.

Comparing the spectra at different angles in Fig. 3(e), it is clear that the spectrum of $\theta=0^{\circ}$ is somewhat lower than those of $\theta=30^{\circ}$ and $\theta=45^{\circ}$
at lower harmonic orders near and below the threshold,
but is remarkably higher than those in the HHG plateau region. Since the yields of below-threshold harmonics for $\theta=0^{\circ}$ are
one order or several orders of magnitude higher than those in the plateau region for intermediate angles of $\theta=30^{\circ}$ and $\theta=45^{\circ}$,
one can expect that EUV pulses with high ellipticity obtained  with the ellipticity hump of $\theta=0^{\circ}$ at lower harmonic orders will also be remarkably
brighter than those obtained with the plateau harmonics of intermediate angles. We will  discuss this point in Fig. 5.

With further increasing the laser ellipticity to $\varepsilon=0.2$, results in the third column of Fig. 3 are somewhat similar to those in the second column,
but the ellipticity of harmonics  in Fig. 3(g) is somewhat more irregular  than in Fig. 3(d), and the intensities of spectra in Fig. 3(h)
are one order of magnitude lower than in Fig. 3(e). The remarkable shift of the dipole minima with the change of the laser ellipticity can also be seen in Fig. 3(i). It should be mentioned that in Figs. 3(h) and 3(i),
the agreement between the spectra and the dipoles for the minima
is not as remarkable as in other columns of Fig. 3, suggesting
that the form of the effective momentum $\textbf{p}'_{sr}$ used here is more applicable for
small ellipticity.In addition, the spectral minima can be influenced by the contributions of excited states to HHG which are not included in SFA.
By comparison, the intersections of the spectra  at different angles match better with those of relevant dipoles in Fig. 3.
Similar phenomena have been discussed in \cite{Chen37,Chen40}.

\begin{figure}[t]
\begin{center}
\rotatebox{0}{\resizebox *{8.5cm}{8.5cm} {\includegraphics {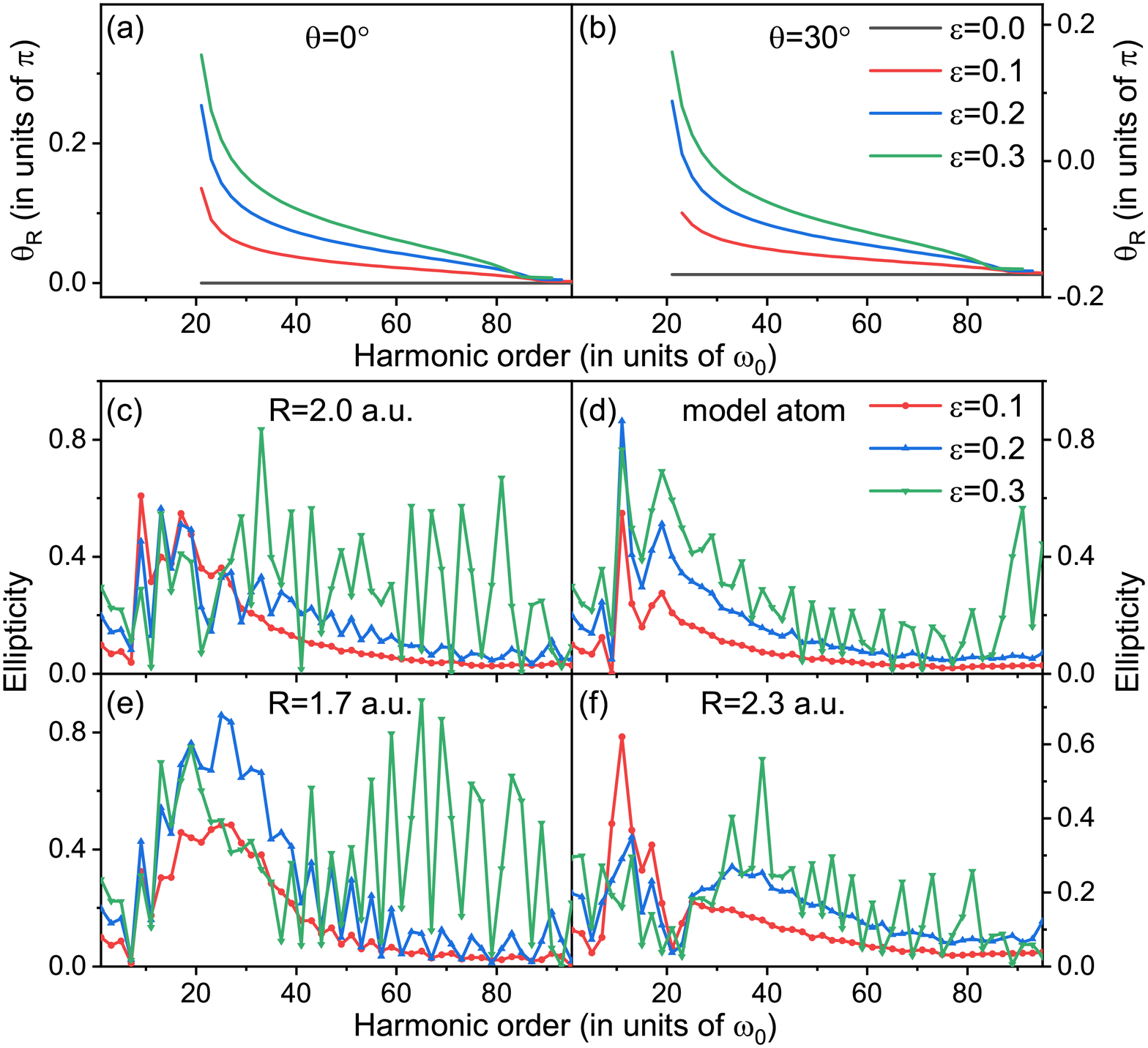}}}
\end{center}
\caption{Recollision angles of HHG short trajectories
from aligned molecules H$_{2}^{+}$ at  $\theta=0^{\circ}$ (a) and $\theta=30^{\circ}$ (b) for different laser ellipticity $\varepsilon$.
In (c)-(f), we show the ellipticity of harmonics for H$_{2}^{+}$ with $R=2$ a.u. (c), $R=1.7$ a.u. (e) and $R=2.3$ a.u. (f) at $\theta=0^{\circ}$ and for model atom with similar $I_p$ to H$_{2}^{+}$ (d).
The laser ellipticity used for obtaining the curves 
is as shown.} \label{Fig. 4}
\end{figure}

From the results in Fig. 3, one can conclude that the large ellipticity of harmonics for a small laser ellipticity is also closely associated with the effect of intramolecular interference,
which plays an important role in the emission of parallel harmonics and has a relatively small role in the perpendicular one, resulting in
a phase difference between parallel and perpendicular harmonics. From the perpendicular spectra of gray curves in the second row of Fig. 3 (also see Fig. 2),
one can observe that the perpendicular spectra do not show the striking hollow structure relating to two-center interference, as seen in the corresponding parallel spectra there.
On the whole, the intensities of perpendicular spectra gradually decrease with the increase of harmonic order.
For the case of parallel alignment of $\theta=0^{\circ}$, this interference between these two atomic centers just occurs at lower harmonic energy
near the threshold, leading to a strong ellipticity of lower harmonic orders.
The influence of two-center interference on ellipticity of HHG from aligned molecules for intermediate angles in a linearly-polarized laser field has been discussed in \cite{Telnov15,Yu19,Dong20,li21}.
Here, we focus on the case of parallel alignment and  elliptically-polarized laser fields.

Next, we further discuss the mechanism of the angle-dependent HHG ellipticity observed in Fig. 3.
Since the interference term $M^r_{s}(\omega,\theta)$ of Eq. (10) in the recombination dipole gives a good description for the dependence of HHG ellipticity on the molecular alignment and the laser ellipticity,
we further analyze the implication of this term.
In Figs. 4(a) and 4(b), we show the recollision angles $\theta_R$, associated with short electron trajectories and defined in Eq. (10),
for  $\theta=0^{\circ}$ and $\theta=30^{\circ}$ at different $\varepsilon$.
For the case of parallel alignment in Fig. 4(a), on the whole, the recollision angle gradually decreases as increasing the harmonic order
for a certain laser ellipticity and  increases with the increase of laser ellipticity for a certain harmonic order.
In particular, for $\varepsilon=0$, the recollision angle is zero and agrees with the alignment angle of $\theta=0^{\circ}$.
For the intermediate angle of $\theta=30^{\circ}$, the situation is different, as shown in Fig. 4(b).
In the case, the absolute value of the recollision angle basically decreases as increasing the laser ellipticity and this decrease is more striking for lower harmonic orders.
According to Eq. (10), in the region of $|\theta_R|\in[0,\pi/2]$, as increasing the absolute value of the angle $\theta_R$, the value of $\cos\theta_R$ decreases,
and the minimal value of the function $M^r_{s}(\omega,\theta)$ will appear at larger $p_{sr}$.
Therefore, for $\theta=0^{\circ}$, the increase of the laser ellipticity which gives rise to the increase of $\theta_R$ will induce
the shift of the  minimum of Eq. (10) towards somewhat larger harmonic energy. This situation reverses for $\theta=30^{\circ}$.
These analyses explain the results in the third row of Fig. 3 and shed light on the spectral and polarization results in other rows of Fig. 3.

\begin{figure}[t]
\begin{center}
\rotatebox{0}{\resizebox *{8.5cm}{6.5cm} {\includegraphics {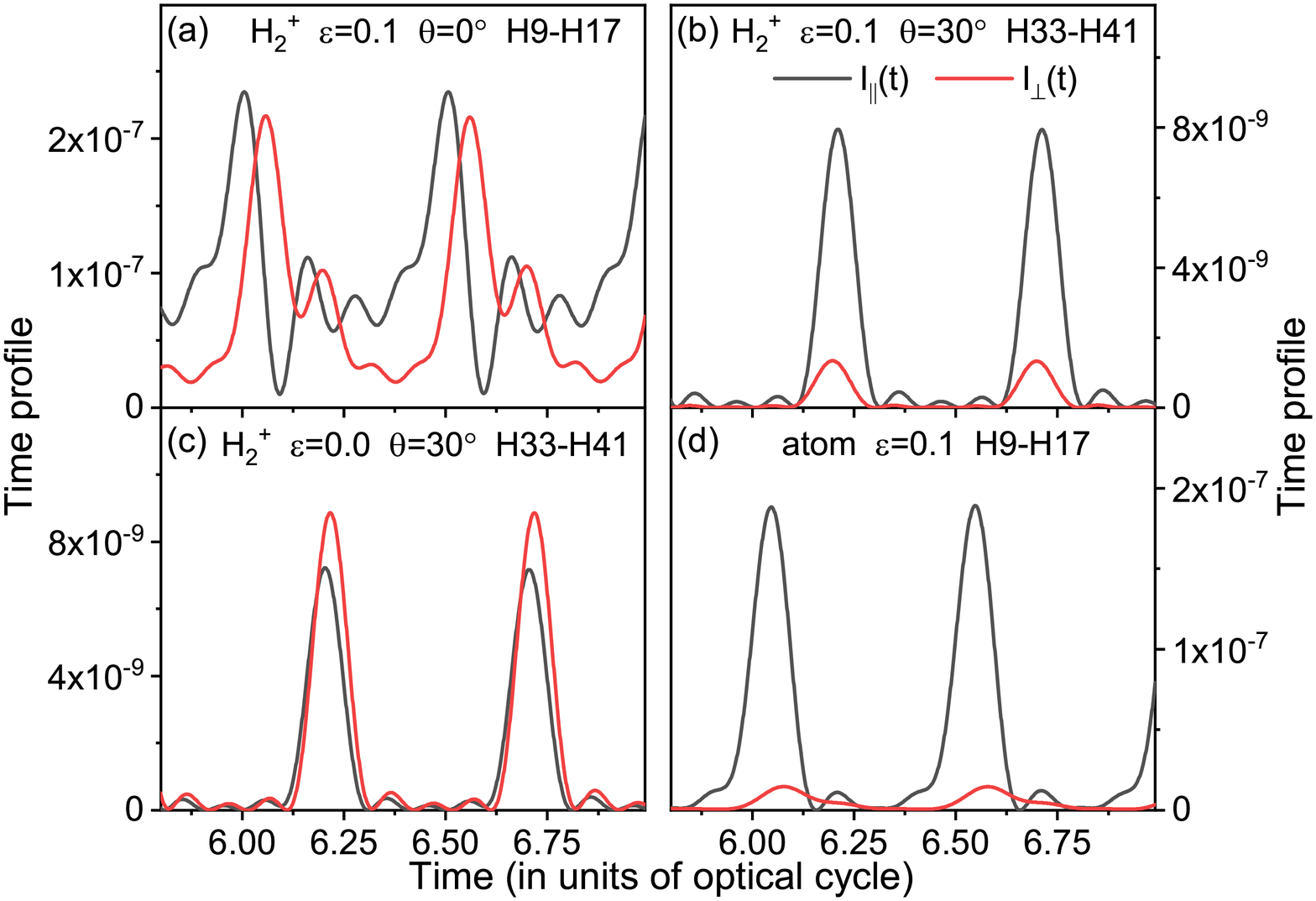}}}
\end{center}
\caption{Trains of pluses $I_{\parallel}(t)$ and $I_{\perp}(t)$ obtained from specific HHG spectral region for aligned molecule H$_{2}^{+}$ at $\theta=0^{\circ}$ (a) and $\theta=30^{\circ}$ (b,c) and for model atom with similar $I_p$ to H$_{2}^{+}$ (d).
In each panel, the laser ellipticity used for obtaining the HHG spectrum and the corresponding specific spectral region are as shown.
} \label{Fig. 5}
\end{figure}

In comparison with intermediate angles, the minimum for $\theta=0^{\circ}$ of H$^+_2$ will appear at lower harmonic energy at which the harmonic spectra usually have larger amplitudes.
This effect also holds for other internuclear distances of H$^+_2$ and for other molecular targets such as N$_2$ with other symmetries. We will discuss the case of N$_2$ later.
Therefore, the parallel alignment of molecules is preferred for obtaining a bright EUV pulse with high ellipticity.
As a comparison, in Figs. 4(c) to 4(f), we show the HHG polarization results of H$^+_2$ at different internuclear distances and laser ellipticity, and
we also show the results of the model atom.
One can observe that the remarkable polarization phenomenon at lower harmonic energy appears in all of the cases.
However, when the results of molecules show an ellipticity hump  around the threshold on the whole, the ellipticity curves of the model atom are somewhat shaper.
From the perspective of shaping a short EUV pulse with a broad energy region of harmonics, the ellipticity hump is also preferred, as shown in Fig. 5.

In Fig. 5, we plot trains of pulses $I_{\parallel}(t)$ and $I_{\perp}(t)$, obtained from the HHG spectra of H$_2^+$ at $\theta=0^{\circ}$.
We also compare the results to cases of  H$_2^+$ at $\theta=30^{\circ}$ and to the model atom.
The expressions of $I_{\parallel}(t)$ and $I_{\perp}(t)$ are as follows \cite{Antoine46,Shi47}:
\begin{equation}
\begin{aligned}
I_{\parallel(\perp)}(t)=\mid\int^{\omega_{u}}_{\omega_{d}} F_{\parallel(\perp)}(\omega)e^{-i\omega t}d\omega\mid^{2}.
\end{aligned}
\end{equation}
Here, $F_{\parallel(\perp)}(\omega)$  are the parallel and perpendicular HHG spectra of Eq. (1).
For H$_2^+$ at $\theta=0^{\circ}$ and for the model atom, we consider the case of  $\varepsilon=0.1$, and
our calculations are performed for the spectral region from H9 to H17 (i.e., $\omega_{d}$ = 9$\omega_{0}$ and $\omega_{u}$ = 17$\omega_{0}$),
at which the harmonics have large ellipticity and high intensities.
For H$_2^+$ at $\theta=30^{\circ}$,  the cases of  $\varepsilon=0$ and $\varepsilon=0.1$ are considered and the spectral region integrated is from H33 to H41.

First, for  H$_2^+$ at $\theta=0^{\circ}$ with $\varepsilon=0.1$, the obtained parallel and perpendicular pulses are comparable for intensities with a remarkable time delay, as seen in Fig. 5(a).
For the case of $\theta=30^{\circ}$ with $\varepsilon=0.1$ in Fig. 5(b),  the pulses of parallel and perpendicular components  show  a small time delay, and the intensities of perpendicular components
are remarkably lower than the parallel ones.
By comparison, for $\varepsilon=0$ in Fig. 5(c), the synthesized parallel and perpendicular pulses show comparable intensities and a small time delay.
Note, for $\varepsilon=0$, the HHG from H$_2^+$ at $\theta=0^{\circ}$ does not show the elliptical-polarization effect due to the absence of perpendicular harmonics.
However, the intensities of the pulses shown in Figs. 5(b) and 5(c) for $\theta=30^{\circ}$ are one order of magnitude lower than those in Fig. 5(a) for $\theta=0^{\circ}$.
The results of the model atom in Fig. 5(d) are somewhat similar to those in Fig. 5(b), with the perpendicular pulse showing small intensities.
The results in Fig. 5 support our previous discussions that harmonics of molecules with parallel alignment
are preferred for obtaining a bright elliptically-polarized EUV pulse.

\begin{figure}[t]
\begin{center}
\rotatebox{0}{\resizebox *{8.5cm}{6.5cm} {\includegraphics {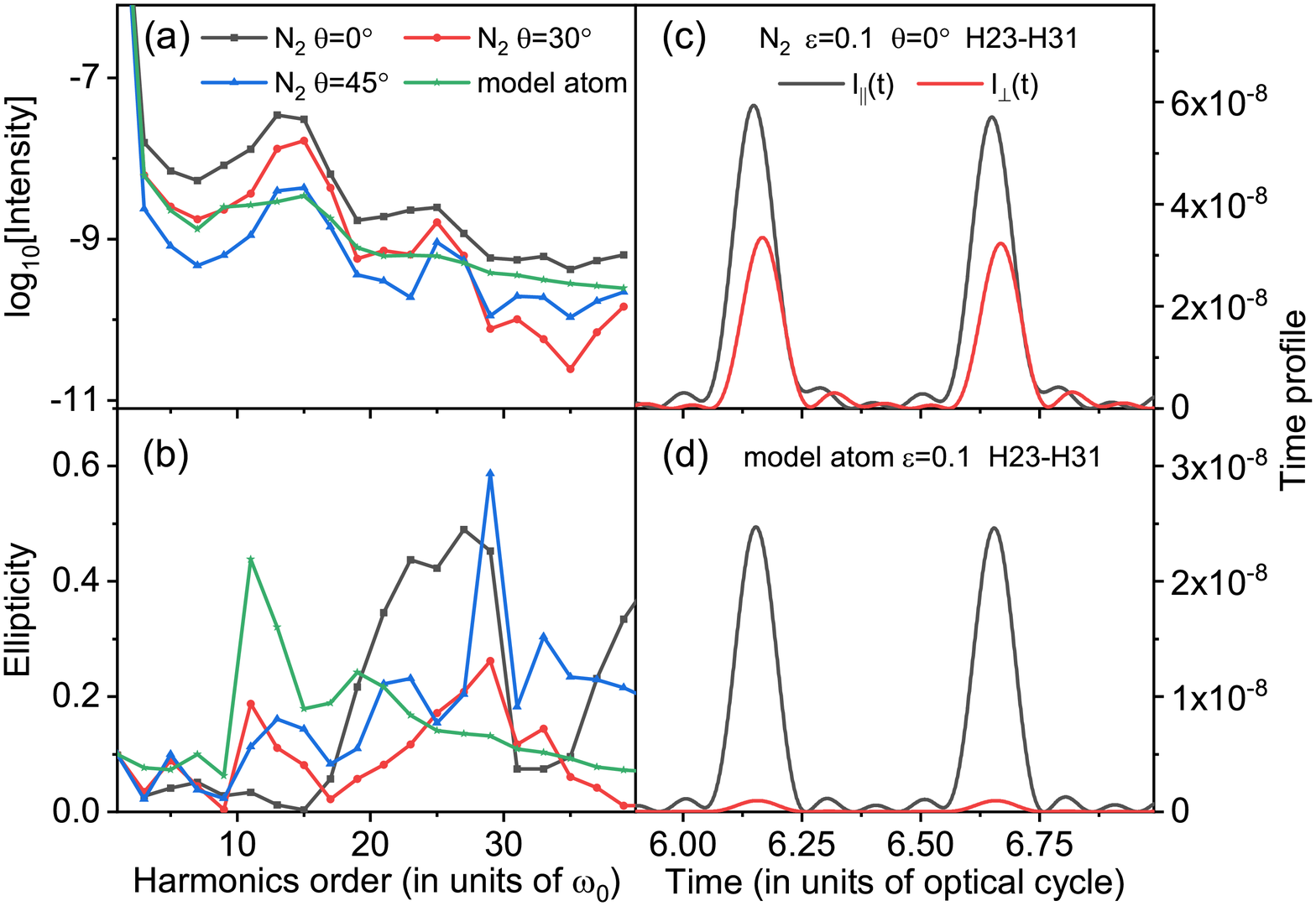}}}
\end{center}
\caption{Spectra of parallel harmonics (a) and ellipticity of harmonics (b) for aligned molecules N$_{2}$ at $\theta=0^{\circ}$ (black square), $\theta=30^{\circ}$ (red circle),
$\theta=45^{\circ}$ (blue triangle), and for model atom with similar $I_p$ to N$_{2}$ (green star), obtained with short-trajectory TDSE simulations.
In (c) and (d), we show trains of pulses $I_{\parallel}$ (black) and $I_{\perp}$ (red) obtained from specific HHG spectral region of H23 to H31 for  N$_{2}$ (c) and the model atom (d).
The laser parameters  used are I$=1.2 \times 10^{14}$/cm$^{2}$ and $\lambda=1400$ nm with $\varepsilon=0.1$.
} \label{Fig. 6}
\end{figure}

To confirm our discussions above, we have also performed calculations for the  N$_{2}$ molecule with 3$\sigma_{g}$ valence orbital,
which can be operated more easily in experiments. To simulate the HHG of N$_2$ with $I_{p}$=0.57 a.u., we have used the model potential $\text{V}(\textbf{r})$  \cite{Wang48} with the form of
$\text{V}(\textbf{r})=-{[(Z-Z_{0}) e^{-\rho \textbf{r}_{1}^{2}}+Z_{0}}]/{\sqrt{\zeta+\textbf{r}_{1}^{2}}}-{[(Z-Z_{0}) e^{-\rho \textbf{r}_{2}^{2}}+Z_{0}}]/{\sqrt{\zeta
+\textbf{r}_{2}^{2}}}$. The expressions of $\textbf{r}^2_{1(2)}$ are as for H$_2^+$.  The relevant parameters used are $Z=5$, $Z_{0}=0.5$, $R=2.079$ a.u., $\zeta=0.5$ and $\rho=1.555$.
Because of the smaller ionization potential  for N$_2$ compared with  H$_{2}^{+}$, here, we use a weaker driving laser intensity of
$I=1.2 \times 10^{14}$/cm$^{2}$ and a longer laser wavelength of $\lambda=1400$ nm.
Relevant results at different angles $\theta$ with $\varepsilon=0.1$ are presented in Fig. 6.
For comparison,  results of a model atom with similar $I_p$ to N$_2$ are also presented here. We focus on lower harmonic orders near the threshold.
One can observe from Fig. 6(a) that the parallel spectra of N$_2$ at $\theta=0^{\circ}$ show the larger intensities than cases of other angles and the model atom.
Around H25, somewhat higher than the threshold harmonic of H19, harmonics of $\theta=0^{\circ}$ also show a ellipticity hump, more remarkable than other cases, as seen in Fig. 6(b).
We mention that in our extended simulations with changing the laser parameters, this hump holds.
The trains of pulses obtained from the spectral region of H23 to H31 for N$_2$ at the parallel alignment and for the model atom are presented in Figs. 6(c) and 6(d).
When the results of model atom for parallel and perpendicular pulses of $I_{\parallel}(t)$ and $I_{\perp}(t)$ do not show a obvious time delay in Fig. 6(d),
the time delay for results of N$_2$ can be clearly identified in Fig. 6(c).
In addition, the relative intensity of the perpendicular pulse in comparison with the parallel one is also remarkably larger for  N$_2$ than for the model atom.

Different from H$_2^+$, whose ground-state wavefunction is dominated by the $1s$ orbital,
the valence-orbital wavefunction of real N$_2$ is composed of the $2p_z$ orbital (about $70\%$) and $1s$ and $2s$ orbitals (about $30\%$) \cite{Kanai31}.
The bound-continuum transition dipole $\textbf{d}(\textbf{p})$ calculated with  $2p_z$-combination molecular orbital has a $\sin$-type interference term,
in contrast to the $\cos$-type one observed in Eq. (8). For $\cos$-type interference term,
the minimum of Eq. (10) appears at the momentum of $p_{sr}=\pi/(R\cos\theta_R)$, and for $\sin$-type one,
that appears at $p_{sr}=2\pi/(R\cos\theta_R)$ which is larger than the $\cos$ one.
Accordingly, the interference minima in the HHG spectra for $\cos$-type molecules generally appear at lower harmonic energy than $\sin$-type ones.
This situation is somewhat more complex for molecules with a mix of  $\cos$-type and  $\sin$-type interference terms such as N$_2$.
However, in both cases, the minima appear at lower harmonic energy for smaller  angles $\theta_R$ (which is near to the alignment angle $\theta$ for a small laser ellipticity)
and harmonics at lower energy near the threshold usually have larger amplitudes.
It is the reason that the parallel alignment is preferred for generating a strong elliptically-polarized EUV pulse, as explored in the paper.

\section{Summary}
In conclusion, we have studied the polarization properties of HHG from aligned molecules in strong elliptically-polarized laser fields with small laser ellipticity.
We have shown that the addition of the vertical component of the laser field with a small intensity has important influences on not only
the yields but also the polarization of harmonics. In particular, for the parallel alignment of the molecule, for which the perpendicular harmonics disappear in a linearly-polarized laser field,
the addition of the small vertical component induces a strong emission of perpendicular harmonics and an accompanying high ellipticity of harmonics near or below the threshold with forming a ellipticity hump.
We show that the phenomenon arises from the effects of two-center interference which influence differently on parallel and perpendicular harmonics, resulting in
a remarkable phase difference between these two harmonic components. As the intensities of parallel and perpendicular harmonics are usually comparable near the threshold,
the inherent phase difference and comparable intensities of these two components  lead to the appearance of the HHG ellipticity hump around the threshold.
By comparison, the HHG ellipticity  for larger alignment angles appears at higher harmonic orders located in the plateau region of the HHG spectra.
Because harmonics near or below the threshold usually have remarkably higher intensities than those in the HHG plateau, the ellipticity hump for the parallel alignment
suggests a manner for generating a bright elliptically-polarized EUV pulses.
Because the intramolecular interference generally occurs at lower harmonic orders for the parallel alignment,
we expect that the ellipticity hump discussed here will also appear for more species of molecules.

\section*{Acknowledgements}
This work was supported by the National Key Research and Development Program of China (Grant No. 2018YFB0504400),
the National Natural Science Foundation of China (Grants No. 91750111 and No. 11904072),
and the Fundamental Research Funds for the Central Universities, China (Grant No. GK201801009).


\begin{thebibliography}{2}
\bibitem{Soc1} A. McPherson, G. Gibson, H. Jara, U. Johann, T. S. Luk, I. A. McIntyre, K. Boyer, and C. K. Rhodes,
Studies of multiphoton production of vacuum-ultraviolet radiation in the rare gases, J. Opt. Soc. Am. B \textbf{4}, 595 (1987).
\bibitem{Huillier2}  A. L'Huillier, K. J. Schafer, and K. C. Kulander, Theoretical aspects of intense field harmonic generation, J. Phys. B \textbf{24}, 3315 (1991).
\bibitem{Tong3} X.-M. Tong and Shih-I Chu, Theoretical study of multiple high-order harmonic generation by intense ultrashort pulsed laser fields:
A new generalized pseudospectral time-dependent method, Chem. Phy. \textbf{217}, 119 (1997).
\bibitem{Lein4} M. Lein, N. Hay, R. Velotta, J. P. Marangos, and P. L. Knignt, Role of the intramoleculaer phase in hign-Order harmonic generation, Phys. Rev. Lett. \textbf{88}, 183903 (2002).
\bibitem{Itatani5} J. Itatani, J. Levesque, D. Zeidler, H. Niikura, H. P\'{e}pin, J. C. Kieffer, P. B. Corkum, and D. M. villeneuve, Tomographic imaging of molecular orbitals, Nature (London) \textbf{432}, 867 (2004).
\bibitem{Corkum6} P. B. Corkum and F. Krausz,  Attosecond science, Nat. Phys, \textbf{3}, 381 (2007).
\bibitem{Vampa7} G. Vampa, C. R. McDonald, G. Orlando, D. D. Klug, P. B. Corkum, and T. Brabec, Theoretical analysis of high-harmonic generation in solids, Phys. Rev. lett. \textbf{113}, 073901 (2014).
\bibitem{Tao8} Z. Tao, C. Chen, T. Szilvasi, M. Keller, M. Mavrikakis, H. Kapteyn, and M. Murnane, Direct time-domain observation of attosecond final-state lifetimes in photoemission from solids, Science \textbf{353}, 62 (2016).
\bibitem{Dejean9} N. T. Dejean, and A. Rubio, Atomic-like high-harmonic generation from two-dimensional materials, Sci. Adv. \textbf{4}, eaao5207 (2018).

\bibitem{Corkum10} P. B. Corkum, Plasma perspective on strong field multiphoton ionization, Phys. Rev. Lett. \textbf{71}, 1994 (1993).
\bibitem{Lewenstein11} M. Lewenstein, Ph. Balcou, M. Yu. Ivanov, A. L'Huillier, and P. B. Corkum,
Theory of high-harmonic generation by low-frequency laser fields, Phys. Rev. A \textbf{49}, 2117 (1994).

\bibitem{Phan12} N. L. Phan, C. T. Le,  V. H. Hoang  and V. H. Le, Odd-even harmonic generation from oriented CO molecules
in linearly polarized laser fields and the influence of the dynamic core-electron polarization, Phys. Chem. Chem. Phys. \textbf{21}, 24177 (2019).


\bibitem{Zhou13} X. Zhou, R. Lock, N. Wagner, W. Li, H. C. Kapteyn, and M. M. Murnane, Elliptically polarized high-order harmonic emission from molecules in linearly polarized laser fields,
Phys. Rev. Lett. \textbf{102}, 073902 (2009).

\bibitem{Levesque14} J. Levesque, Y. Mairesse, N. Dudovich, H. P\'{e}pin, J.-C. Kieffer, P. B. Corkum, and D. M. Villeneuve,
Polarization state of high-order harmonic emission from aligned molecules, Phys. Rev. Lett. \textbf{99}, 243001 (2007).
\bibitem{Telnov15} S.-K. Son, D. A. Telnov, and Shih-I Chu,
Probing the origin of elliptical high-order harmonic generation from aligned molecules in linearly polarized laser field, Phys. Rev. A \textbf{82}, 043829 (2010).
\bibitem{Seideman16} S. Ramakrishna, P. A. J. Sherratt, A. D. Dutoi, and T. Seideman,
Origin and implication of ellipticity in high-order harmonic generation from aligned molecules, Phys. Rev. A \textbf{81}, 021802(R) (2010).
\bibitem{Sherratt17} P. A. J. Sherratt, S. Ramakrishna, and T. Seideman,
Signatures of the molecular potential in the ellipticity of high-order harmonics from aligned molecules, Phys. Rev. A \textbf{83}, 053425 (2011).
\bibitem{Strelkov18} V. V. Strelkov, A. A. Gonoskov, I. A. Gonoskov, and M. Yu. Ryabikin, Origin for ellipticity of high-order harmonics generated
in atomic gases and the sublaser-cycle evolution of harmonic polarization, Phys. Rev. Lett. \textbf{107}, 043902 (2011).

\bibitem{Yu19} S. J. Yu, B. Zhang, Y. P. Li, S. P. Yang, and Y. J. Chen,
Ellipticity of odd-even harmonics from oriented asymmetric molecules in strong linearly polarized laser fields, Phys. Rev. A \textbf{90}, 053844 (2014).
\bibitem{Dong20} F. L. Dong, Y. Q. Tian, S. J. Yu, S. Wang, S. P. Yang, and Y. J. Chen,
Polarization properties of below-threshold harmonics from aligned molecules H$_{2}^{+}$ in linearly polarized laser fields, Opt. Express \textbf{23}, 18106 (2015).
\bibitem{li21} W. Y. Li, F. L. Dong, S. J. Yu, S. Wang, S. P. Yang, and Y. J. Chen,
Ellipticity of near-threshold harmonics from stretched molecules, Opt. Express \textbf{23}, 31010 (2015).
\bibitem{Budil22} K. S. Budil, P. Sali\`{e}res, A. L'Huillier, T. Ditmire, and M. D. Perry, Influence of ellipticity on harmonic generation, Phys. Rev. A \textbf{48}, R3437 (1993).
\bibitem{Dietrich23} P. Dietrich, N. H. Burnett, M. Ivanov,  and P. B. Corkum,
High-harmonic generation and correlated two-electron multiphoton ionization with elliptically polarized light, Phys. Rev. A \textbf{50}, R3585 (1994).
\bibitem{Antoine24} P. Antoine, B. Carr\'{e}, A. L'Huillier, and M. Lewenstein, Polarization of high-order harmonics, Phys. Rev. A \textbf{55}, 1314 (1997).
\bibitem{Strelkov25} V. V. Strelkov, Theory of high-order harmonic generation and attosecond pulse emission by a low-frequency elliptically polarized laser field, Phys. Rev. A \textbf{74}, 013405 (2006).

\bibitem{Strelkov26} V. V. Strelkov, M. A. Khokhlova, A. A. Gonoskov, I. A. Gonoskov, and M. Yu. Ryabikin,
High-order harmonic generation by atoms in an elliptically polarized laser field: Harmonic polarization properties and laser threshold ellipticity, Phys. Rev. A \textbf{86}, 013404 (2012).
\bibitem{Sarantseva27} M. V. Frolov, N. L. Manakov, T. S. Sarantseva, and A. F. Starace,
High-order-harmonic-generation spectroscopy with an elliptically polarized laser field, Phys. Rev. A \textbf{86}, 063406 (2012).
\bibitem{Frolov28} T. S. Sarantseva, A. A. Silaev and N. L. Manakov,
High-order-harmonic generation in an elliptically polarized laser field: analytic form of the electron wave packet, J. Phys. B \textbf{50}, 074002 (2017).


\bibitem{Velotta29} R. Velotta, N. Hay, M. B. Mason, M. Castillejo, and J. P. Marangos, High-order harmonic generation in aligned molecules, Phys. Rev. Lett. \textbf{87}, 183901 (2001).
\bibitem{Litvinyuk30} I. V. Litvinyuk, K. F. Lee, P. W. Dooley, D. M. Rayner, D. M. Villeneuve, and P. B. Corkum, Alignment-dependent strong field ionization of molecules, Phys. Rev. Lett. \textbf{90}, 233003 (2003).


\bibitem{Kanai31} T. Kanai, S. Minemoto, and H. Sakai, Ellipticity dependence of high-order harmonic generation from aligned molecules, Phys. Rev. Lett. \textbf{98}, 053002 (2007).

\bibitem{Mairesse32} Y. Mairesse, N. Dudovich, J. Levesque, M. Yu. Ivanov, P. B. Corkum, and D. M. Villeneuve,
Electron wavepacket control with elliptically polarized laser light in high harmonic generation from aligned molecules, New. J. Phys. \textbf{10}, 025015 (2008).

\bibitem{Odzak33} S. Od\v{z}ak and D. B. Milo\v{s}evi\'{c}, Role of ellipticity in high-order harmonic generation by homonuclear diatomic molecules, Phys. Rev. A \textbf{82}, 023412 (2010).


\bibitem{Yang34} H. Yang, P. Liu, R. X. Li and Z. Z. Xu, Ellipticity dependence of the near-threshold harmonics of H$_{2}$ in an elliptical strong laser field, Opt. Express \textbf{21}, 28676 (2013).

\bibitem{Becker35} Y. Q. Xia, and A. Jaro\'{n}-Becker,
Multielectron contributions in elliptically polarized high-order harmonic emission from nitrogen molecules, Opt. Lett. \textbf{39}, 1461 (2014).




\bibitem{Feit36} M. D. Feit, J. A. Fleck, Jr., and A. Steiger, Solution of the Schr\"{o}dinger equation by a spectral method, J. Comput. Phys. \textbf{47}, 412 (1982).

\bibitem{Chen37} Y. J. Chen, J. Liu, and B. Hu, Reading molecular messages from high-order harmonic spectra at different orientation angles, J. Chem. Phys. \textbf{130}, 044311 (2009).


\bibitem{Salieres38} P. Sali\`{e}res, B. Carr\'{e}, L. Le D\'{e}roff, F. Grasbon, G. G. Paulus, H. Walther, R. Kopold, W. Becker,
D. B. Milo\v{s}evi\'{c}, A. Sanpera, and M. Lewenstein, Feynman's path-integral approach for intense-laser-atom interactions, Science \textbf{292}, 902 (2001).


\bibitem{Lewenstein39} M. Lewenstein, P. Sali\`{e}res, and A. L'Huillier, Phase of the atomic polarization in high-order harmonic generation, Phys. Rev. A \textbf{52}, 4747 (1995).


\bibitem{Chen40} Y. J. Chen and B. Hu, Role of ionization in orientation dependence of molecular high-order harmonic generation, J. Chem. Phys. \textbf{131}, 244109 (2009).
\bibitem{Xie41} X. J. Xie, S. J. Yu, W. Y. Li, S. Wang, and Y. J. Chen, Routes of odd-even harmonic emission from oriented polar molecules, Opt. Express \textbf{26}, 18578 (2018).

\bibitem{Chen42} Y. J. Chen, and B. Hu, Intense field ionization of diatomic molecules: Two-center interference and tunneling, Phys. Rev. A \textbf{81}, 013411 (2010).
\bibitem{Li43} W. Y. Li, S. Wang, Y. Z. Shi, S. P. Yang, and Y. J. Chen, Probing the structure of stretched molecular ions with high-harmonic spectroscopy, J. Phys. B \textbf{50}, 085003 (2017).



\bibitem{Chen44} C. Chen, D. X. Ren, X. Han, S. P. Yang, and Y. J. Chen, Time-resolved harmonic emission from aligned molecules in orthogonal two-color fields, Phy. Rev. A \textbf{98}, 063425 (2018).
\bibitem{Tong45} X.-M. Tong, and Shih-I Chu, Probing the spectral and temporal structures of high-order harmonic generation in intense laser pulses, Phys. Rev. A \textbf{61}, 021802(R) (2000).





\bibitem{Antoine46} P. Antoine, A. L. Huillier, and M. Lewenstein, Attosecond pulse trains using high-order harmonics, Phys. Rev. Lett. \textbf{77}, 1234 (1996).
\bibitem{Shi47} Y. Z. Shi, S. Wang, F. L. Dong, Y. P. Li, and Y. J. Chen, Classical effect for enhanced high harmonic yield in ultrashort laser pulses with a moderate laser intensity, J. Phys. B \textbf{50}, 065004 (2017).



\bibitem{Wang48} S. Wang, J. Cai, and Y. J. Chen, Ionization dynamics of polar molecules in strong elliptical laser fields,  Phy. Rev. A \textbf{96}, 043413 (2017).

\bibitem [*] {2} chenchao1202@163.com
\bibitem [\dag] {1} chenyjhb@gmail.com
\end{thebibliography}
\end{document}